\documentclass[aps,prl,twocolumn,superscriptaddress,tightenlines,showpacs,preprintnumbers,amsmath,amssymb]{revtex4}

\usepackage{graphicx} 
\usepackage{dcolumn}  
\usepackage{bm}

\def\Journal#1#2#3#4{{#1} {\bf #2}, #3 (#4)}

\def\NIMA{Nucl. Instr. and Meth. A}

\def\PLB{{Phys. Lett.}  B}
\def\PRL{Phys. Rev. Lett.}
\def\PRD{{Phys. Rev.} D}

\def\be{\begin{equation}}
\def\ee{\end{equation}}
\def\bea{\begin{eqnarray}}
\def\eea{\end{eqnarray}}

\def\nbb{275 million $B\bar{B}$ pairs}
\def\dsup{D_{\rm sup}}
\def\dfav{D_{\rm fav}}
\def\dpisig{6.4\sigma}
\def\dksig{2.3\sigma}
\def\rb{r_B}
\def\rblimit{0.27}
\def\rd{r_D}
\def\deltab{\delta_B}
\def\deltad{\delta_D}
\def\stat{{\rm stat}}
\def\syst{{\rm syst}}
\def\LR{{\cal R}}

\begin{document}

\preprint{
  \vbox{
  }
}

\title{
  \quad\\[0.5cm]  
  \boldmath 
  Study of the Suppressed Decays 
  $B^- \to [K^+\pi^-]_D K^-$ and $B^- \to [K^+\pi^-]_D \pi^-$
}

\affiliation{Budker Institute of Nuclear Physics, Novosibirsk}
\affiliation{Chiba University, Chiba}
\affiliation{Chonnam National University, Kwangju}
\affiliation{University of Cincinnati, Cincinnati, Ohio 45221}
\affiliation{Gyeongsang National University, Chinju}
\affiliation{University of Hawaii, Honolulu, Hawaii 96822}
\affiliation{High Energy Accelerator Research Organization (KEK), Tsukuba}
\affiliation{Hiroshima Institute of Technology, Hiroshima}
\affiliation{Institute of High Energy Physics, Chinese Academy of Sciences, Beijing}
\affiliation{Institute of High Energy Physics, Vienna}
\affiliation{Institute for Theoretical and Experimental Physics, Moscow}
\affiliation{J. Stefan Institute, Ljubljana}
\affiliation{Kanagawa University, Yokohama}
\affiliation{Korea University, Seoul}
\affiliation{Kyungpook National University, Taegu}
\affiliation{Swiss Federal Institute of Technology of Lausanne, EPFL, Lausanne}
\affiliation{University of Ljubljana, Ljubljana}
\affiliation{University of Maribor, Maribor}
\affiliation{University of Melbourne, Victoria}
\affiliation{Nagoya University, Nagoya}
\affiliation{Nara Women's University, Nara}
\affiliation{National Central University, Chung-li}
\affiliation{National United University, Miao Li}
\affiliation{Department of Physics, National Taiwan University, Taipei}
\affiliation{H. Niewodniczanski Institute of Nuclear Physics, Krakow}
\affiliation{Nihon Dental College, Niigata}
\affiliation{Niigata University, Niigata}
\affiliation{Osaka City University, Osaka}
\affiliation{Osaka University, Osaka}
\affiliation{Panjab University, Chandigarh}
\affiliation{Peking University, Beijing}
\affiliation{Princeton University, Princeton, New Jersey 08545}
\affiliation{Saga University, Saga}
\affiliation{University of Science and Technology of China, Hefei}
\affiliation{Seoul National University, Seoul}
\affiliation{Sungkyunkwan University, Suwon}
\affiliation{University of Sydney, Sydney NSW}
\affiliation{Tata Institute of Fundamental Research, Bombay}
\affiliation{Toho University, Funabashi}
\affiliation{Tohoku Gakuin University, Tagajo}
\affiliation{Tohoku University, Sendai}
\affiliation{Department of Physics, University of Tokyo, Tokyo}
\affiliation{Tokyo Institute of Technology, Tokyo}
\affiliation{Tokyo Metropolitan University, Tokyo}
\affiliation{Tokyo University of Agriculture and Technology, Tokyo}
\affiliation{University of Tsukuba, Tsukuba}
\affiliation{Virginia Polytechnic Institute and State University, Blacksburg, Virginia 24061}
\affiliation{Yonsei University, Seoul}
  \author{M.~Saigo}\affiliation{Tohoku University, Sendai} 
  \author{K.~Abe}\affiliation{High Energy Accelerator Research Organization (KEK), Tsukuba} 
  \author{K.~Abe}\affiliation{Tohoku Gakuin University, Tagajo} 
  \author{H.~Aihara}\affiliation{Department of Physics, University of Tokyo, Tokyo} 
  \author{M.~Akatsu}\affiliation{Nagoya University, Nagoya} 
  \author{Y.~Asano}\affiliation{University of Tsukuba, Tsukuba} 
  \author{V.~Aulchenko}\affiliation{Budker Institute of Nuclear Physics, Novosibirsk} 
  \author{T.~Aushev}\affiliation{Institute for Theoretical and Experimental Physics, Moscow} 
  \author{S.~Bahinipati}\affiliation{University of Cincinnati, Cincinnati, Ohio 45221} 
  \author{A.~M.~Bakich}\affiliation{University of Sydney, Sydney NSW} 
% ...
\collaboration{The Belle Collaboration}

\date{\today}
\noaffiliation

\begin{abstract}
  We report a study of the suppressed decays $B^- \to [K^+\pi^-]_D K^-$
  and $B^- \to [K^+\pi^-]_D \pi^-$,
  where $[K^+\pi^-]_D$ indicates that 
  the $K^+\pi^-$ pair originates from a neutral $D$ meson. 
  These decay modes are sensitive to the Unitarity Triangle angle $\phi_3$.
  We use a data sample containing \nbb\ recorded 
  at the $\Upsilon(4S)$ resonance with the Belle detector at the 
  KEKB asymmetric $e^+e^-$ storage ring. 
  The signal for $B^- \to [K^+\pi^-]_D K^-$ 
  is not statistically significant,
  and we set a limit
  $\rb < \rblimit$ at $90\%$ confidence level,
  where $\rb$ is the magnitude of the ratio of amplitudes
  $\left| A(B^- \to \bar{D}^0K^-) / A(B^- \to D^0K^-) \right|$.
  We observe a signal with $\dpisig$ statistical significance 
  in the related mode, $B^- \to [K^+\pi^-]_D \pi^-$.
\end{abstract}

\pacs{11.30.Er, 12.15.Hh, 13.25.Hw, 14.40.Nd}

\maketitle

\begin{figure}
  \begin{center}
    \includegraphics[width=0.40\textwidth]{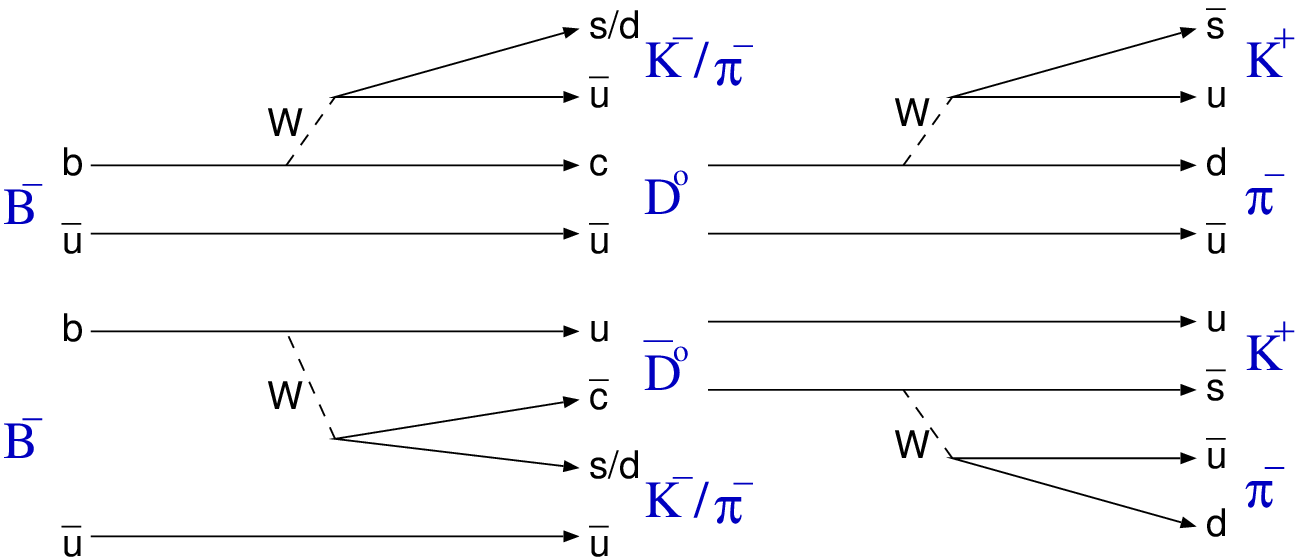}
    \caption{$B^-\to[K^+\pi^-]_D K^-$ and $B^-\to[K^+\pi^-]_D \pi^-$ decays.}
    \label{fig:btodcsk}
  \end{center}
\end{figure}

Precise measurements of the elements of the 
Cabibbo-Kobayashi-Maskawa matrix~\cite{km} 
constrain the standard model and may reveal new physics.
However, the extraction of the Unitarity Triangle angle $\phi_3$
is a challenging measurement even with modern high luminosity $B$ factories. 
Several methods for measuring $\phi_3$ use the interference 
between $B^- \to D^0K^-$ and $B^- \to \bar{D}^0K^-$, 
which occurs when $D^0$ and $\bar{D}^0$ decay 
to common final states~\cite{bs,glw}. 
As noted by Atwood, Dunietz and Soni (ADS)~\cite{ads},
$CP$ violation effects are enhanced if the final state is chosen so
that the interfering amplitudes have comparable magnitudes;
the archetype uses $B^- \to [K^+\pi^-]_D K^-$,
where $[K^+\pi^-]_{D}$ indicates that the $K^+\pi^-$ pair 
originates from a neutral $D$ meson. 
In this case, the color-allowed $B$ decay 
followed by the doubly Cabibbo-suppressed $D$ decay 
interferes with  the color-suppressed $B$ decay 
followed by the Cabibbo-allowed $D$ decay (Fig.~\ref{fig:btodcsk}). 
Previous studies of this decay mode have not found any signals~\cite{ba}.
For the suppressed decay $B^- \to [K^+\pi^-]_D \pi^-$,
both topology and phenomenology are similar to $B^- \to [K^+\pi^-]_D K^-$. 

In this analysis, 
the favored decays 
$B^- \to [K^-\pi^+]_D h^-$, where $h = \pi$ or $K$,
are used as control samples to reduce systematic uncertainties. 
The same selection criteria for the suppressed decay modes 
are applied to the control samples whenever possible. 
Throughout this report, 
charge conjugate states are implied except where explicitly mentioned 
and we denote the analyzed decay modes as follows:
\begin{equation}\nonumber
  \begin{array}{ccc}
    \textrm{Suppressed decay} & B^- \to [K^+\pi^-]_D h^- & B^- \to \dsup h^- \\
    \textrm{Favored decay}    & B^- \to [K^-\pi^+]_D h^- & B^- \to \dfav h^- \\
  \end{array}
\end{equation}

The results are based on a data sample containing \nbb,
collected with the Belle detector at the KEKB asymmetric energy $e^+e^-$ 
collider~\cite{kekb} operating at the $\Upsilon(4S)$ resonance. 
The Belle detector is a large-solid-angle magnetic
spectrometer that consists of a silicon vertex detector (SVD),
a 50-layer central drift chamber (CDC), 
an array of aerogel threshold \v{C}erenkov counters (ACC),
a barrel-like arrangement of time-of-flight scintillation counters (TOF), 
and an electromagnetic calorimeter (ECL) comprised of CsI(Tl) crystals 
located inside a superconducting solenoid coil 
that provides a $1.5 \ {\rm T}$ magnetic field.  
An iron flux-return located outside of the coil is instrumented 
to detect $K_L^0$ mesons and to identify muons (KLM).  
The detector is described in detail elsewhere~\cite{Belle}.
Two different inner detector configurations were used. 
For the first sample of 152 million $B\bar{B}$ pairs, 
a 2.0 cm radius beampipe and a 3-layer silicon vertex detector were used;
for the latter 123 million $B\bar{B}$ pairs,
a 1.5 cm radius beampipe, a 4-layer silicon detector
and a small-cell inner drift chamber were used~\cite{Ushiroda}.

Neutral $D$ mesons are reconstructed by combining 
two oppositely charged tracks. 
For each track, information from ACC, TOF and specific ionization 
measurements from the CDC are used to determine a $K/\pi$ likelihood ratio 
$P(K/\pi) = {\cal L}_K/({\cal L}_K + {\cal L}_\pi)$, 
where ${\cal L}_K$ and ${\cal L}_\pi$ are kaon and pion likelihoods. 
We use the particle identification requirements 
$P(K/\pi) > 0.8$ for kaons and $P(K/\pi) < 0.2$ for pions.
These requirements select kaons (pions) with momentum dependent 
efficiencies of $80 - 95\%$ ($90 - 95\%$)
and pion (kaon) misidentification probabilities of $5 - 20\%$ ($15 - 20\%$).
$D$ candidates are required to have an invariant mass 
within $\pm2.5\sigma$ of the nominal $D$ mass: 
$1.850 \ {\rm GeV}/c^2 < M(K\pi) < 1.879 \ {\rm GeV}/c^2$. 
To improve the momentum determination, 
tracks from the $D$ candidate are refitted 
according to the nominal $D$ mass hypothesis 
and the reconstructed vertex position (a mass-and-vertex-constrained fit).
$B$ mesons are reconstructed by combining $D$ candidates 
with primary charged hadron candidates. 
The signal is identified by two kinematic variables, 
the energy difference 
$\Delta E = E_D  + E_{h^-} - E_{\rm beam}$ 
and the beam-energy-constrained mass 
$M_{\rm bc} = \sqrt{E^{2}_{\rm beam} - (\vec{p}_D  + \vec{p}_{h^-})^{2}}$,
where $E_D$ is the energy of the $D$ candidate, 
$E_{h^-}$ is the energy of the $h^-$, 
and $E_{\rm beam}$ is the beam energy, 
all evaluated in the center-of-mass (cm) frame;
$\vec{p}_D$ and $\vec{p}_{h^-}$ are 
the momenta of the $D$ and $h^-$ in the cm frame. 
Event candidates are accepted if they have 
$5.2 \ {\rm GeV}/c^2 < M_{\rm bc} < 5.3 \ {\rm GeV}/c^2$ 
and $|\Delta E| < 0.2 \ {\rm GeV}$. 
If there is more than one candidate in an event,
we select the best candidate on the basis of a $\chi^{2}$ 
determined from the difference between the 
measured and nominal values of $M(K\pi)$ and $M_{\rm bc}$.  

To suppress the large background from the two-jet-like 
$e^+e^- \to q\bar{q} \ (q = u, d, s, c)$ continuum processes, 
variables that characterize the event topology are used. 
We construct a Fisher discriminant~\cite{fisher} of Fox-Wolfram moments 
called the Super-Fox-Wolfram ($SFW$)~\cite{sfw}, 
where the Fisher coefficients are optimized by maximizing 
the separation between $B\bar{B}$ events and continuum events. 
Furthermore, $\cos \theta_B$, the cosine of the angle in the cm system 
between the $B$ flight direction with respect to the beam axis 
is also used to distinguish $B\bar{B}$ events 
from continuum events. 
These two independent variables, $SFW$ and $\cos \theta_B$, 
are combined to form a likelihood ratio 
$\LR = {\cal L}_{\rm sig}/({\cal L}_{\rm sig} + {\cal L}_{\rm cont})$, 
where ${\cal L}_{\rm sig}$ and ${\cal L}_{\rm cont}$ 
are likelihoods calculated from the $SFW$ and $\cos \theta_B$ distributions 
of signal and continuum background events, respectively. 
We optimize the $\LR$ requirement by maximizing 
$S/\sqrt{S+N}$, 
where $S$ and $N$ denote the expected numbers 
of signal and background events in the signal region. 
For $B^- \to \dsup K^-$ ($B^- \to \dsup \pi^-$) 
we require $\LR > 0.85$ ($\LR > 0.75$),
which retains $44.8\%$ ($57.6\%$) of the signal events 
and removes $96.2\%$ ($93.2\%$) of the continuum background. 

For $B^- \to \dsup K^-$, 
there can be a contribution from $B^- \to D^0\pi^-$, $D^0 \to K^+K^-$, 
which has the same final state and can peak under the signal. 
In order to reject these events, 
we veto events that satisfy 
$1.843 \ {\rm GeV}/c^2 < M(KK) < 1.894 \ {\rm GeV}/c^2$. 
The favored decay $B^- \to \dfav h^-$ 
can also cause a peaking background for the suppressed decay modes 
if both the kaon and pion from the $\dfav$ decay are misidentified.
Therefore, we veto events for which the invariant mass of the $K\pi$ pair 
is inside the $D$ mass window when the mass assignments are exchanged. 
After applying this veto, 
the residual background from $K\pi$ misidentification is found to be negligible.
The three-body charmless decay $B^- \to K^+K^-\pi^-$
can peak inside the signal regions of $\Delta E$ and $M_{\rm bc}$ for 
$B^- \to \dsup K^-$.
This background is estimated from the $\Delta E$ distribution
of events in a $D$ mass sideband, defined as 
$1.637 \ {\rm GeV}/c^2 < M(K\pi) < 1.836 \ {\rm GeV}/c^2$ and 
$1.893 \ {\rm GeV}/c^2 < M(K\pi) < 2.093 \ {\rm GeV}/c^2$. 
The estimated peaking background inside the $\Delta E$ signal region
is $1.7 \pm 0.9$ events, which we subtract
from the observed $B^- \to \dsup K^-$ yield.
As a check, we estimate the expected background level from the measured 
$B^- \to K^+K^-\pi^-$ branching fraction~\cite{hhh}.
Using this result, 
the expected background in our signal region is $2.1 \pm 0.6$ events, 
assuming that the $B^- \to K^+K^-\pi^-$ yield is 
uniformly distributed in phase space.
For $B^- \to \dsup \pi^-$, 
the peaking background estimated from the $D$ mass sideband 
is consistent with zero. 

\begin{table*}
  \caption{
    Summary of the results. 
    For the $B^- \to \dsup K^-$ signal yield, 
    the peaking background contribution has been subtracted. 
    The first two errors on the measured production branching fractions
    are statistical and systematic, respectively,
    and the third is due to the uncertainty in the $B^- \to \dfav h^-$
    product branching fraction used for normalization.
  }
  \begin{ruledtabular}
    \begin{tabular}{ l | c c c c c c}
      Mode & Product branching & Efficiency & Signal Yield & Statistical  & Measured product & Upper limit \\
           & fraction from~\cite{pdg} & ($\%$) & & significance  & branching fraction & ($90\%$C.L.)\\
      \hline
      $B^- \to \dsup K^-$ & $-$                     & $12.9 \pm 0.2$ & 
      $   8.5\,^{ +6.0}_{ -5.3}$ & $\dksig$  & $(3.2\,^{+2.2}_{-2.0} \pm 0.2 \pm 0.5) \times 10^{-7}$ & $6.3 \times 10^{-7}$ \\
      $B^- \to \dsup \pi^-$ & $(6.9 \pm 0.7) \times 10^{-7}$ & $20.1 \pm 0.2$ & 
      $  28.5\,^{ +8.1}_{ -7.4}$ & $\dpisig$ & $(6.6\,^{+1.9}_{-1.7} \pm 0.4 \pm 0.3) \times 10^{-7}$ & $-$ \\
      $B^- \to \dfav K^-$ & $(1.4 \pm 0.2) \times 10^{-5}$ & $12.9 \pm 0.2$ & 
      $ 376.0\,^{+21.8}_{-21.1}$ & $-$       & $-$ & $-$ \\
      $B^- \to \dfav \pi^-$ & $(1.9 \pm 0.1) \times 10^{-4}$ & $20.3 \pm 0.2$ & 
      $8181.9\,^{+94.0}_{-93.3}$ & $-$       & $-$ & $-$ \\
    \end{tabular}
  \end{ruledtabular}
  \label{tab:yield}
\end{table*}

The signal yields are extracted 
using binned maximum likelihood fits to the $\Delta E$ distributions
of events in the $M_{\rm bc}$ signal region, 
($5.27 \ {\rm GeV}/c^2 < M_{\rm bc} < 5.29 \ {\rm GeV}/c^2$).
Backgrounds from decays such as $B^- \to D\rho^-$ and $B^- \to D^{*}\pi^-$ 
are distributed in the negative $\Delta E$ region 
and make a small contribution to the signal region. 
The shape of this $B\bar{B}$ background is modeled 
with a smoothed histogram obtained from generic Monte Carlo (MC) samples. 
The continuum background populates the entire $\Delta E$ region. 
The shape of the continuum background is modeled with a linear function. 
The slope is determined from the $\Delta E$ distribution 
of the $M_{\rm bc}$ sideband 
($5.20 \ {\rm GeV}/c^2 < M_{\rm bc} < 5.26 \ {\rm GeV}/c^2$). 
The $\Delta E$ fitting function is the sum of two Gaussians for the signal, 
the linear function for the continuum, 
and the smoothed histogram for the $B\bar{B}$ background distribution. 

\begin{figure}
  \begin{center}
    \includegraphics[width=0.4\textwidth]{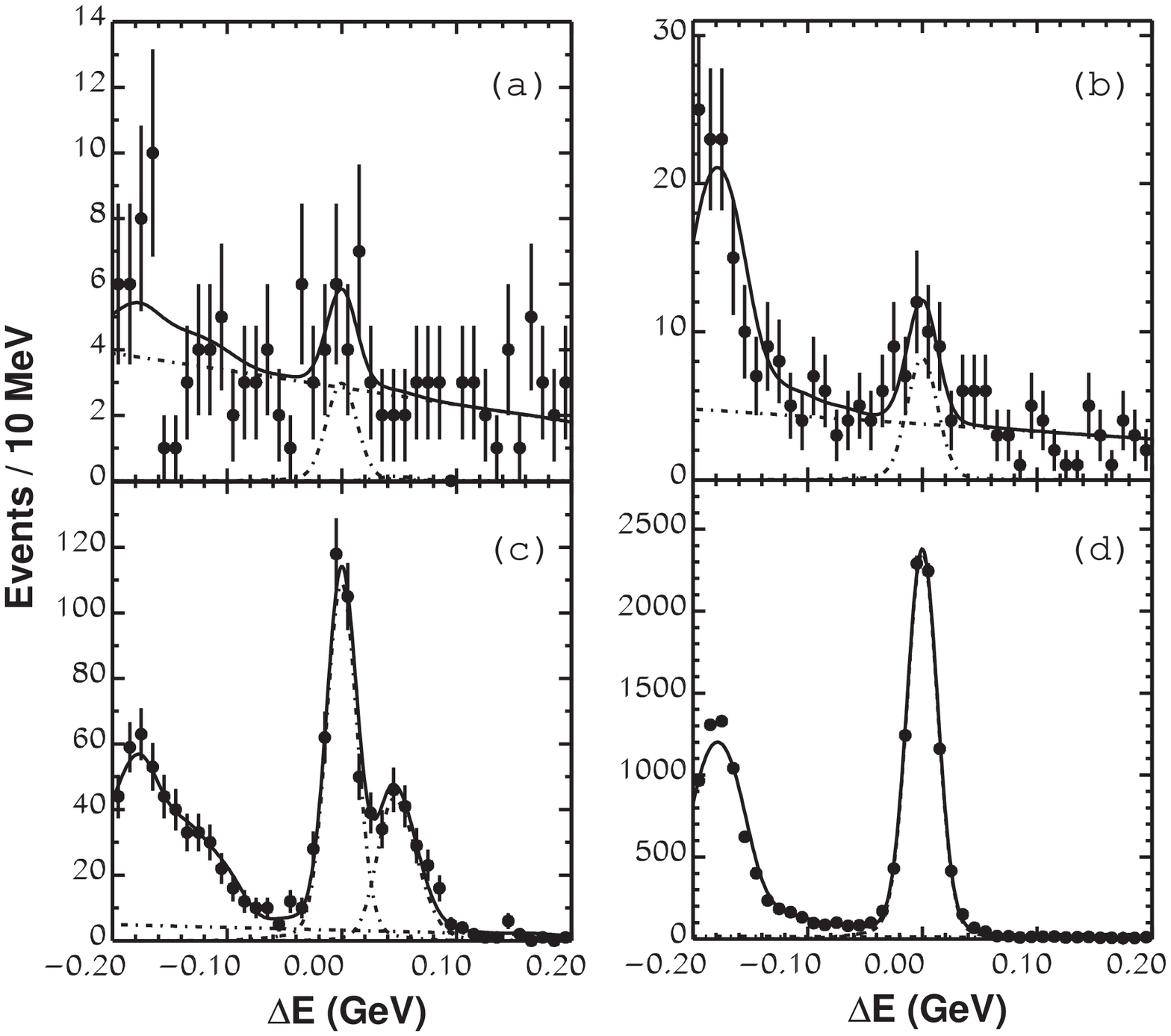}
    \caption{
      $\Delta E$ fit results for 
      (a) $B^- \to \dsup K^-$, 
      (b) $B^- \to \dsup \pi^-$, 
      (c) $B^- \to \dfav K^-$, and 
      (d) $B^- \to \dfav \pi^-$. 
      Charge conjugate modes are included in these plots.
    }
    \label{fig:fitting}
  \end{center}
\end{figure}

In the fit to the $\Delta E$ distribution of $B^- \to \dfav \pi^-$, 
the free parameters are the position, width and area of the signal peak, 
and the normalizations of continuum and $B\bar{B}$ backgrounds. 
For the signal, 
the relative widths and areas of the two Gaussians
are fixed from the signal MC. 
For the $B^- \to \dfav K^-$ fit, 
the position and width of the signal peak are fixed 
from the $B^- \to \dfav \pi^-$ fit results. 
To fit the feed-across from $B^- \to \dfav \pi^-$, 
we use a Gaussian shape where the left and right sides 
of the peak have different widths since the shift 
caused by wrong mass assignment makes the shape asymmetric. 
The shape parameters of this function are fixed to values 
determined by the fit to the $B^- \to \dfav \pi^-$ distribution 
using a kaon mass hypothesis for the prompt pion. 
The areas of signal and feed-across from $B^- \to D\pi^-$, 
and the normalizations of continuum and $B\bar{B}$ backgrounds 
are floated in the fit. 
For $B^- \to \dsup K^-$ and $B^- \to \dsup \pi^-$, 
the signal and $B\bar{B}$ background shapes 
are modeled using the fit results for the 
$B^- \to \dfav K^-$ and $B^- \to \dfav \pi^-$ modes, respectively.  
The area of the feed-across from $\dsup \pi^-$ 
is estimated as the measured yield of $B^- \to \dsup \pi^-$ 
multiplied by the $\pi$ to $K$ misidentification probability. 
However, the areas of the signal and the normalizations of 
continuum and $B\bar{B}$ backgrounds are floated.  
The fit results are shown in Fig.~\ref{fig:fitting}. 
The numbers of events for $B^- \to \dsup h^-$ and $\dfav h^-$, 
and the statistical significances of the $B^- \to \dsup h^-$ signals 
are given in Table~\ref{tab:yield}. 
The statistical significance is defined as 
$\sqrt{-2\ln({\cal L}_{0}/{\cal L}_{\rm max})}$, 
where ${\cal L}_{\rm max}$ is the maximum likelihood in the $\Delta E$ fit 
and ${\cal L}_{0}$ is the likelihood 
when the signal yield is constrained to be zero. 
The uncertainty in the peaking background contribution 
is taken into account in the statistical significance calculation. 

The ratio of branching fractions is defined as
\begin{equation} \nonumber 
  R_{Dh} \equiv 
  \frac{{\cal B}(B^-\to \dsup h^-)}{{\cal B}(B^-\to \dfav h^-)} = 
  \frac{N_{\dsup h^-}/\epsilon_{\dsup h^-}}{N_{\dfav h^-}/\epsilon_{\dfav h^-}}, 
\end{equation}
where $N_{\dsup h}$ ($N_{\dfav h}$) and  $\epsilon_{\dsup h^-}$ ($\epsilon_{\dfav h^-}$)
are the number of signal events and the reconstruction efficiency for the decay
$B^- \to \dsup h^-$ ($B^- \to \dfav h^-$),
and are given in Table~\ref{tab:yield}.

The ratios $R_{Dh}$ are calculated to be
\begin{eqnarray}
  R_{DK} &=& ( 2.3 \, ^{+1.6}_{-1.4} (\stat) \pm 0.1 (\syst) ) \times 10^{-2}, 
  \nonumber \\
  R_{D\pi} &=& ( 3.5 \, ^{+1.0}_{-0.9} (\stat) \pm 0.2 (\syst) ) \times 10^{-3}.
  \nonumber
\end{eqnarray}
Since the signal for $B^- \to \dsup K^-$ is not significant,
we set an upper limit at the $90\%$ confidence level (C.L.) of 
$R_{DK} < 4.4 \times 10^{-2}$,
where we take the likelihood function as a single Gaussian 
with width given by the quadratic sum of statistical and systematic errors,
and the area is normalized in the physical region 
of positive branching fraction. 

Most of the systematic uncertainties from the detection efficiencies 
and the particle identification cancel when taking the ratios, 
since the kinematics of the $B^- \to \dsup h^-$ 
and $B^- \to \dfav h^-$ processes are similar. 
The systematic errors are due to 
uncertainties in the yield extraction ($4.7\%-5.4\%$) 
and the efficiency difference between 
$B^- \to \dsup h^-$ and $B^- \to \dfav h^-$ ($1.3\%-1.7\%$). 
The uncertainties in the signal shapes and the $q\bar q$ background shapes 
are determined by varying the shape of the fitting function by $\pm1\sigma$. 
The uncertainties in the $B\bar B$ background shapes are determined 
by fitting the $\Delta E$ distribution in the region 
$-0.07 \ {\rm GeV} < \Delta E < 0.20 \ {\rm GeV}$ 
ignoring the $B\bar{B}$ background contributions. 
The uncertainties in the efficiency differences are determined using signal MC.
The total systematic error is the sum in quadrature of the above uncertainties.

The product branching fractions for $B^- \to \dsup h^-$ are determined as
\begin{equation} \nonumber
  {\cal B}(B^- \to \dsup h^-) = {\cal B}(B^- \to \dfav h^-) \times R_{Dh},
\end{equation}
and are given in Table~\ref{tab:yield}. 
A third uncertainty arises due to the error in the branching fraction
of $B^- \to \dfav h^-$, which is taken from~\cite{pdg}.
The uncertainties are statistics-dominated. 
For the $B^- \to \dsup K^-$ branching fraction, 
we set an upper limit at the $90\%$ C.L. of 
${\cal B}(B^- \to \dsup K^-) < 6.3 \times 10^{-7}$. 
For $B^- \to \dsup \pi^-$,
our measured branching fraction is consistent with expectation 
neglecting the contribution from $B^- \to \bar{D}^0 \pi^-$.

The ratio $R_{DK}$ is related to $\phi_3$ by
\begin{eqnarray}
  R_{DK} = \rb^2 + \rd^2 + 2 \rb \rd \cos \phi_3 \cos \delta, \nonumber
\end{eqnarray}
where~\cite{pdg}
\begin{eqnarray}
  \rb & \equiv & \left| \frac{A(B^- \to \bar{D}^0K^-)}{A(B^- \to D^0K^-)} \right|, 
  \:\:\:\:\: 
  \delta \equiv \deltab + \deltad, 
  \nonumber \\
  \rd & \equiv & \left| \frac{A(D^0 \to K^+\pi^-)}{A(D^0 \to K^-\pi^+)} \right| = 
  0.060 \pm 0.003, \nonumber
\end{eqnarray}
and $\deltab$ and $\deltad$ are the strong phase differences 
between the two $B$ and $D$ decay amplitudes, respectively. 
Using the above result, we obtain a limit on $\rb$. 
The least restrictive limit is obtained 
allowing $\pm 1\sigma$ variation on $\rd$ and assuming maximal interference
($\phi_3 = 0^\circ, \delta = 180^\circ$ or $\phi_3 = 180^\circ, \delta = 0^\circ$) 
and is found to be $\rb < \rblimit$.

\begin{figure}
  \begin{center}
    \includegraphics[width=0.4\textwidth]{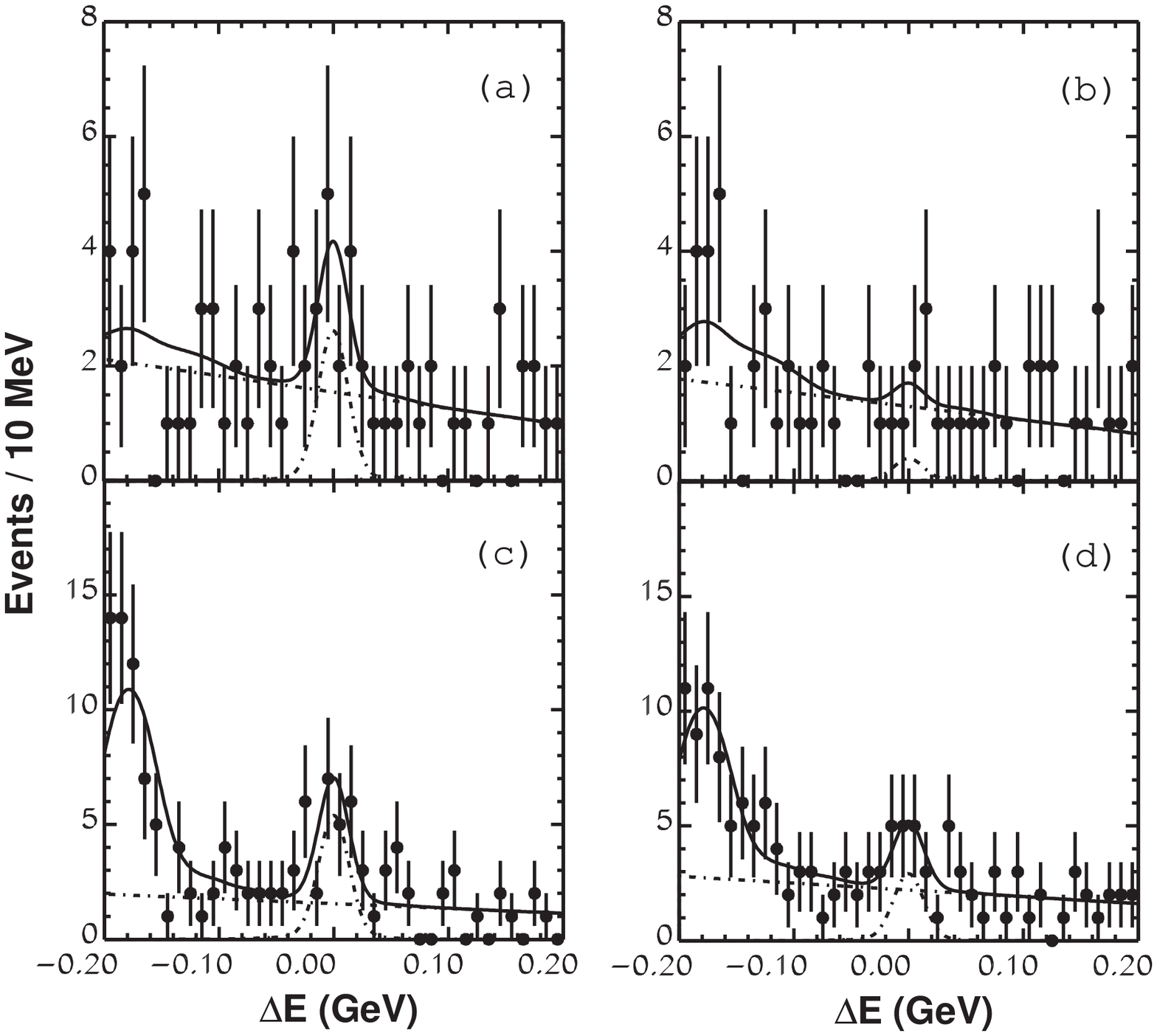}
    \caption{
      $\Delta E$ fit results for 
      (a) $B^- \to \dsup K^-$, 
      (b) $B^+ \to \dsup K^+$, 
      (c) $B^- \to \dsup \pi^-$, and 
      (d) $B^+ \to \dsup \pi^+$.
    }
    \label{fig:acp}
  \end{center}
\end{figure}

\begin{table}
  \caption{
    Signal yields and partial rate asymmetries.
  }
  \begin{ruledtabular}
    \begin{tabular}{ l | c c c c c}
      Mode & $N(B^-)$ & & $N(B^+)$ & & ${\cal A}_{Dh}$\\
      \hline
      $B \to \dsup K$ & $ 8.2\,^{+5.0}_{-4.3}$ & & $ 0.5\,^{+3.5}_{-2.8}$ & & $0.88\,^{+0.77}_{-0.62} \pm 0.06$ \\
      $B \to \dsup \pi$ & $18.8\,^{+6.3}_{-5.5}$ & & $10.1\,^{+5.5}_{-4.8}$ & & $0.30\,^{+0.29}_{-0.25} \pm 0.06$ \\
    \end{tabular}
  \end{ruledtabular}
  \label{tab:acp}
\end{table}

We search for partial rate asymmetries 
${\cal A}_{Dh}$ in $B^\mp \to \dsup h^\mp$ decay, 
fitting the $B^-$ and $B^+$ yields separately for each mode, 
where ${\cal A}_{Dh}$ is determined as
\begin{equation}\nonumber 
  {\cal A}_{Dh} \equiv 
  \frac{
    {\cal B}(B^- \to \dsup h^-) - {\cal B}(B^+ \to \dsup h^+)
  }{
    {\cal B}(B^- \to \dsup h^-) + {\cal B}(B^+ \to \dsup h^+)
  }.
\end{equation}
The peaking background for $B^\mp \to \dsup K^\mp$ 
is subtracted assuming no $CP$ asymmetry. 
The fit results are shown in Fig.~\ref{fig:acp} and Table~\ref{tab:acp}. 
We find
\begin{eqnarray}
  {\cal A}_{DK} & = & 0.88 \, ^{+0.77}_{-0.62} (\stat) \pm 0.06 (\syst),
  \nonumber \\
  {\cal A}_{D\pi} & = & 0.30 \, ^{+0.29}_{-0.25} (\stat) \pm 0.06 (\syst), 
  \nonumber 
\end{eqnarray}
where the systematic uncertainties arise from 
possible biases in the analysis algorithms 
(estimated from the $B^\mp \to \dfav \pi^\mp$ control sample to be $2.5\%$); 
uncertainties in the extraction of the $B^-$ and $B^+$ yields 
(estimated by varying fitting parameters by $\pm1\sigma$ to be $4.9\%$); 
asymmetry in the particle identification efficiency of prompt kaons 
(estimated in~\cite{acp_kpi} to be $0.6\%$).
We assume no $CP$ asymmetry in the peaking background,
and do not assign any systematic uncertainty from this source~\cite{bkgd_asp}.

In summary, using \nbb\ collected with the Belle detector, 
we report studies of the suppressed decay $B^-\to \dsup h^-$ ($h = K,\pi$). 
We observe $B^- \to \dsup \pi^-$ for the first time, 
with a significance of $\dpisig$. 
The size of the signal is consistent with 
expectation based on measured branching fractions~\cite{pdg}. 
The significance for $B^- \to \dsup K^-$ is $\dksig$ 
and we set an upper limit on the ratio of $B$ decay amplitudes 
$\rb < \rblimit$ at $90\%$ confidence level.
This result is consistent with previous searches~\cite{ba},
and with the measurement of $\rb$ 
in the decay $B^- \to DK^-$, $D \to K_S^0\pi^+\pi^-$~\cite{da},
which provides the most precise current determination of $\phi_3$.

We thank the KEKB group for the excellent operation of the
accelerator, the KEK Cryogenics group for the efficient
operation of the solenoid, and the KEK computer group and
the NII for valuable computing and Super-SINET network
support.  We acknowledge support from MEXT and JSPS (Japan);
ARC and DEST (Australia); NSFC (contract No.~10175071,
China); DST (India); the BK21 program of MOEHRD and the CHEP
SRC program of KOSEF (Korea); KBN (contract No.~2P03B 01324,
Poland); MIST (Russia); MESS (Slovenia); NSC and MOE
(Taiwan); and DOE (USA).


\begin{thebibliography}{99}
\bibitem{bkgd_asp}
  An assumption of $30\%$ $CP$ asymmetry in the peaking background
  would lead to a shift of $5.5\%$ in ${\cal A}_{DK}$.
\end{thebibliography}
\end{document}